\documentstyle[12pt,epsf]{article}

\begin{document}

\title{\Large\bf Vicious Random Walkers and a Discretization of 
Gaussian Random Matrix Ensembles}
\author{Taro Nagao and Peter J. Forrester$^{\dagger}$}
\date{}
\maketitle

\begin{center}
\it
Department of Physics, Graduate School of Science, Osaka University, \\
Toyonaka, Osaka 560-0043, Japan
\end{center}
\begin{center}
\it
 $^{\dagger}$Department of Mathematics and Statistics, University of Melbourne, \\
Victoria 3010, Australia
\end{center}

\bigskip
\begin{center}
\bf Abstract 
\end{center}
\par
\bigskip
\noindent
The vicious random walker problem on a
one dimensional lattice is considered. Many walkers 
take simultaneous 
steps on the lattice and the configurations in which two of them 
arrive at the same site are prohibited. It is known that 
the probability distribution of $N$ walkers after $M$ steps can be written 
in a determinant form. Using an integration technique borrowed from the 
theory of random matrices, we show that arbitrary $k$-th order correlation 
functions of the walkers can be expressed as quaternion determinants whose 
elements are compactly expressed in terms of symmetric Hahn polynomials. 

\newpage
\noindent
\section{Introduction}
\setcounter{equation}{0}
\renewcommand{\theequation}{1.\arabic{equation}}

The vicious walker problem first introduced by Fisher\cite{MEF} 
and then developed by Forrester\cite{PJF} (see also \cite{ZLE}) 
recently attracts much attention in mathematical physics. Fascinating 
connections to other research fields, such as Young tableaux 
in combinatorics\cite{KJ}, 
Kardar-Parisi-Zhang (KPZ) universality in the theory of growth 
process\cite{PS} and the theory of random matrices\cite{JB}, have been 
revealed one after another. In the context of random 
matrix theory, the ensemble of vicious walkers in one dimension 
corresponds to a discretization of the Gaussian ensembles of 
random matrices. Therefore the theory of discretized random 
matrices is expected to shed light on all of the related problems. 
Indeed discretized random matrices in the guise of discrete Coulomb
gases are central to the work of Johansson\cite{KJ,KJ1,KJ2} in these
directions.
\par
\begin{figure}
\epsfxsize=8cm
\centerline{\epsfbox{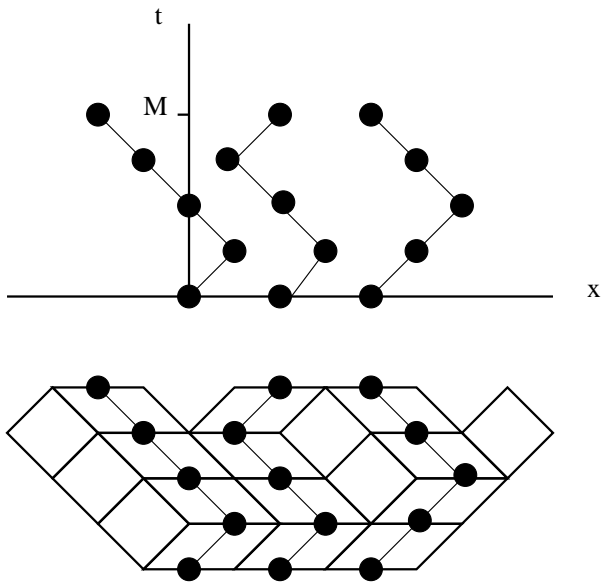}}
\caption{\label{f1} Nonintersecting paths representing a vicious walker
configuration with $N = 3$ walkers and $M=4$ steps. In the top diagram
these paths are drawn in the $x$--$t$ plane. In the bottom diagram
superimposed on the paths is the equivalent rhombi tiling, involving
left sloping rhombi (step to the left), right sloping rhombi
(step to the right) and vertical rhombi.}
\end{figure}

Suppose that there are $N$ walkers on a one dimensional lattice. 
In the lock step version of the model
(as distinct from the random turns version considered in the recent
work\cite{FoR}),
at each time step each walker moves to the left or right 
one lattice site with 
equal probability.  Walkers are "vicious" so that two or 
more walkers are prohibited to arrive at the same site simultaneously.       
The $j$-th walker starts at the position $x = 2 j - 2$ and, 
after $M$ steps, arrives at $x = X_j$. 
The walker configurations form nonintersecting paths in the $x$--$t$ plane.
An example is given in Figure \ref{f1}. Furthermore it has long been
realized \cite{BH82} that there is a one-to-one correspondence between
such nonintersecting paths and random rhombus tilings of a hexagon, or
suitable truncation thereof, involving three types of rhombi. This is
also illustrated in Figure \ref{f1}.

The number of lock step 
paths is known to be expressed as a binomial determinant\cite{EG,GOV,KGV} 
\begin{equation}
P_1(X_1,X_2,\cdots,X_N) = \det\left[ \left( \begin{array}{c} M \\ \displaystyle 
\frac{M + X_j}{2} - l + 1 \end{array} \right) \right]_{j,l = 1,2,\cdots,N}.
\end{equation}
The binomial determinant can be further rewritten as a product formula
\begin{equation}
P_1(X_1,X_2,\cdots,X_N) = 2^{-N(N-1)/2} \prod_{j=1}^N \frac{(M + N - j)!}{
\displaystyle \left( \frac{M + X_j}{2} \right)! \left( \frac{M - X_j}{2} 
+ N - 1 \right)!} \prod_{j>l}^N (X_j - X_l), 
\end{equation}
where $X_1 < X_2 < \cdots < X_N$. 
Now we introduce new variables
\begin{eqnarray}
x_j & = & \frac{X_j - N + 1}{2}, \nonumber \\ 
L & = & M + N - 1
\end{eqnarray}
in order to obtain a compact and symmetric expression
\begin{equation}\label{1.4}
P_1(x_1,x_2,\cdots,x_N) = C_{MN} \prod_{j=1}^N \sqrt{w(x_j)} 
\prod_{j>l}^N \mid x_j - x_l \mid.
\end{equation}
Here
\begin{equation}
C_{MN} = \prod_{j=1}^N (M + N - j)!
\end{equation}
and
\begin{equation}\label{1.6}
w(x) = \frac{1}{\displaystyle 
\left[ \left( \frac{L}{2} + x \right)! 
\left( \frac{L}{2} - x \right)! \right]^2}.
\end{equation}

\vspace{.5cm}
\par
In some applications of the vicious walker problem one imposes the 
additional constraint that each walker returns to the initial 
position after $2 M$ steps. 
One example is the rhombus tiling problem of Figure \ref{f1} with the
region to be tiled extended to be symmetrical about $t=M$ and thus
made into a hexagon\cite{KJ2}.
In such cases the number of paths 
is given by the square of $P_1(x_1,x_2,\cdots,x_N)$:
\begin{equation}\label{1.7}
P_2(x_1,x_2,\cdots,x_N) = C_{MN}^2 \prod_{j=1}^N w(x_j)
\prod_{j>l}^N \mid x_j - x_l 
\mid^2.
\end{equation}
As will be revised below, the function $w(x)$ is a special case of the
Hahn weight function from the theory of discrete orthogonal polynomials.
The probability density (\ref{1.7}) with the Hahn weight 
is intimately related to Hahn polynomials and so has been termed the
Hahn ensemble\cite{KJ1,KJ2}.
\par
We are interested in the number of paths under the condition that 
$k$ walkers take fixed positions $x_1,x_2,\cdots,x_k$ after $M$ steps 
($k \leq N$). This number is given by the 
correlation functions 
\begin{equation}\label{1.8}
I^{(\beta)}_k(x_1,x_2,\cdots,x_k) 
= \frac{1}{(N-k)!} \sum_{x_{k+1}=-\infty}^{\infty} 
\sum_{x_{k+2}=-\infty}^{\infty} \cdots  
\sum_{x_N=-\infty}^{\infty} P_{\beta}(x_1,x_2,\cdots,x_N)
\end{equation}
in both the cases $\beta =1$ and $\beta = 2$. 
\par
In the case with returning walkers ($\beta = 2$), 
the evaluation of the correlation 
functions is relatively easy. Introducing monic orthogonal 
polynomials $C_j(x)=x^j + \cdots$ with orthogonality relations 
\begin{equation}
\sum_{x=-\infty}^{\infty} w(x) C_j(x) C_l(x) = \delta_{jl} h_j,
\end{equation}
we can readily obtain 
\begin{equation}
I^{(2)}_k(x_1,x_2,\cdots,x_k) = C_{MN}^2 \prod_{j=0}^{N-1} h_j 
\det[K(x_j,x_l)]_{j,l=1,2,\cdots,k} 
\end{equation}
with
\begin{eqnarray}\label{1.11}
K(x,y) & = & 
\sqrt{w(x) w(y)} \sum_{j=0}^{N-1} \frac{1}{h_j} C_j(x) C_j(y) \nonumber \\ 
& = & \sqrt{w(x) w(y)} \frac{1}{h_{N-1}} \frac{
C_N(x) C_{N-1}(y)- C_{N-1}(x) C_N(y)}{x-y}.
\end{eqnarray}
We call $C_j(x)$ symmetric Hahn polynomials and define them in terms 
of the Hahn polynomials in \S 2. 
\par
The main purpose of this paper is to derive an analogous 
formula for the correlation functions with no returning constraint 
($\beta = 1$). For that purpose, we again make use of 
the above symmetric Hahn polynomials to rewrite the 
correlation functions in the form of a determinant although now with 
 quaternion elements. The Hahn polynomials and their 
symmetrization are introduced in \S 2. In \S 3, we introduce quaternion 
determinant formulas for the correlation functions $I^{(1)}_k(x_1,\cdots,x_k)$. 
In \S 4, the continuous limit to the Gaussian ensembles is discussed.

\section{Symmetric Hahn Polynomials}
\setcounter{equation}{0}
\renewcommand{\theequation}{2.\arabic{equation}}
The Hahn polynomials $Q_n(x)$ are orthogonal polynomials 
with respect to the weight function\cite{RK} 
\begin{equation}\label{2.1}
w_H(x) = \frac{(a + x)!}{a! x!} \frac{(L + 
b - x)!}{b! (L-x)!}
\end{equation}
on a discrete measure ($x$ is restricted to be an integer). 
When the coefficient of the highest order term is fixed as 
\begin{equation}
Q_n(x) = \frac{(n + a + b + 1)_n}{(a + 1)_n (- L)_n} x^n + \cdots,
\end{equation}
we have the orthogonality relation
\begin{equation}
\sum_{x=0}^L w_H(x) Q_m(x) Q_n(x) = H_n \delta_{mn},
\end{equation}
where
\begin{equation}
H_n = \frac{(-1)^n n! (b + 1)_n (n + a + b + 1)_{L+1}}{
L! (2 n + a + b + 1) (-L)_n (a + 1)_n}.
\end{equation}
Here $(a)_n = \Gamma(a + n)/\Gamma(a)$. 
Moreover there is a dual orthogonality relation (completeness relation)
\begin{equation}
\sum_{n=0}^L \frac{1}{H_n} Q_n(x) Q_n(y) = \frac{\delta_{xy}}{w_H(x)},
\end{equation}
which yields the orthogonality relation of the dual Hahn polynomials. 
\par
The weight (\ref{1.6}) is the special case $a=b=-L-1$ of (\ref{2.1}) with
shift of coordinate $x \mapsto x + L/2$. The monic orthogonal polynomials
with respect to (\ref{1.6}), which we have denoted $C_n(x)$ and termed the
symmetric Hahn polynomials, are therefore given in terms of the Hahn
polynomials $Q_n(x)$ by
\begin{equation}
C_n(x) =  \left. \frac{(a + 1)_n (-L)_n}{(n + a + b + 1)_n} 
Q_n\left( x + \frac{L}{2} \right) 
\right|_{a=b = - L - 1}
\end{equation}
and satisfy the orthogonality relation
\begin{equation}\label{2.7}
\sum_{x=-\infty}^{\infty} w(x) C_j(x) C_l(x) = \delta_{jl} h_j
\end{equation}
with
\begin{equation}\label{2.8}
h_j = \frac{j! (2 L - 2 j + 1)! (2 L - 2 j)!}{(2 L - j + 1)! 
[(L - j)!]^4} .
\end{equation}
The corresponding dual orthogonality relation reads 
\begin{equation}\label{2.9}
\sum_{n=0}^L \frac{1}{h_n} C_n(x) C_n(y) = \frac{\delta_{xy}}{w(x)}.
\end{equation}
Since the weight function $w(x)$ defined in (\ref{1.6}) 
is symmetric ($w(-x) = w(x)$), 
$C_n(x)$ are even or odd polynomials corresponding to the 
parity of $n$ ($C_n(-x) = (-1)^n C_n(x)$).
\par
In Ref.\cite{WK}, recurrence relations for the Hahn polynomials are presented. 
In the symmetric limit $a=b = - L - 1$, we find
\begin{equation}
C_{n + 1}(x) = x C_n(x) - \omega_n C_{n-1}(x) 
\end{equation}
and
\begin{equation}
- \left( x + \frac{L}{2} \right)^2 [ C_n(x) - C_n(x-1) ] = 
\alpha_n C_{n+1}(x) + \beta_n C_n(x) + \gamma_n C_{n-1}(x),
\end{equation}
where
\begin{equation}
\omega_n = \frac{n (2 L - n + 2) (L - n + 1)^2}{4 (2 L - 2 n + 3) 
(2 L - 2 n + 1)},
\end{equation}
\begin{equation}
\alpha_n = - n, \ \ \beta_n = - \frac{1}{2} n (2 L - n + 1)
\end{equation}
and
\begin{equation}
\gamma_n = - \frac{n (2 L - n + 1) (2 L - n + 2) 
(L - n + 1)^2}{4 (2 L - 2 n + 3) (2 L - 2 n + 1)}.
\end{equation}

\section{Correlation Functions}
\setcounter{equation}{0}
\renewcommand{\theequation}{3.\arabic{equation}}

In the theory of random real symmetric matrices, 
it is known that the correlation functions among the 
eigenvalues are expressed as quaternion 
determinants\cite{DYSON,MAHOUX,MBOOK}. Although in most cases the integration 
method has been applied to continuous eigenvalue distributions, 
several authors have made attempts to evaluate the correlation 
functions on discrete measures\cite{GAUDIN,MM,SUTH,NW1}. Here we can employ a similar 
procedure to rewrite the multiple sum (\ref{1.8}) in a quaternion 
determinant form. 

\subsection{Quaternion Determinant}

As the first step let us introduce the quaternion determinant. 
A quaternion is defined as a linear combination of 
four basic units $\{1, e_1, e_2, e_3 \}$:
\begin{equation} 
q=q_0+{\bf q} \cdot {\bf e}=q_0+q_1e_1+q_2e_2+q_3e_3, 
\end{equation}
where $q_0, q_1, q_2$ and $q_3$ are real or complex numbers. 
The first part $q_1$ is called the scalar part of $q$. 
The multiplication laws of the four basic units are given by 
\begin{equation} 
1 \cdot 1=1,\;\; 1 \cdot e_j=e_j \cdot 
1=e_j,\;\; j=1,2,3, \nonumber 
\end{equation}
\begin{equation} 
e_1^2=e_2^2=e_3^2=e_1e_2e_3=-1. 
\end{equation} 
Note that the quaternion multiplication is associative 
but in general not commutative. 
The dual ${\hat q}$ of a quaternion $q$ is defined as 
\begin{equation} 
{\hat q} = q_0-{\bf q} \cdot {\bf e}. 
\end{equation} 
A matrix $Q$ with quaternion elements $q_{jl}$ has a dual 
matrix ${\hat Q}=[{\hat q}_{lj}]$. The quaternion units are 
represented as $2 \times 2$ matrices
\begin{displaymath} 
1 \rightarrow 
\left[ \begin{array}{cc} 1 & 0 \\ 0 & 1 \end{array} 
\right], \ \ e_1 \rightarrow \left[ \begin{array}{cc} 0 & -1 \\ 1 & 0 \end{array} \right],
\end{displaymath}
\begin{equation} 
e_2 \rightarrow \left[ \begin{array}{cc} 0 & -i \\ -i & 0 \end{array} \right], 
\ \ e_3 \rightarrow \left[ \begin{array}{cc} i & 0 \\ 0 & -i \end{array} \right], 
\end{equation} 
so that any $2 \times 2$ matrix with complex elements can be identified 
as a $2 \times 2$ representation of a quaternion.
\par
We define a quaternion determinant {\rm Tdet} of a self dual 
$Q$ ( i.e., $Q={\hat Q}$ ) as
\begin{equation} {\rm Tdet}\, Q = \sum_P (-1)^{N-l} \prod_1^l (q_{ab} 
q_{bc} \cdots q_{da})_0.
\end{equation}
Here $P$ denotes any permutation of the indices 
$(1,2,\cdots,N)$ consisting of $l$ exclusive cycles of the form 
$(a \rightarrow b \rightarrow c \rightarrow \cdots \rightarrow d 
\rightarrow a)$ and $(-1)^{N-l}$ is the parity of $P$. The subscript 
$0$ has a meaning that we take the scalar part of the product over 
each cycle.

\subsection{Skew Orthogonal Polynomials}

Next we need to define skew orthogonal polynomials in order 
to give the quaternion determinant expressions for the 
correlation functions. The Schmidt's orthogonalization 
procedure enables us to construct 
monic polynomials $R_n(x)$ of degree $n$ which satisfy 
\begin{displaymath} \langle R_{2m}(x), R_{2n+1}(y) \rangle = 
- \langle R_{2n+1}(x), R_{2m}(y) \rangle = r_m \delta_{mn}, \end{displaymath}
\begin{equation} \langle R_{2m}(x), R_{2n}(y) \rangle = 0, \quad 
 \langle R_{2m+1}(x), R_{2n+1}(y) \rangle = 0,\label{3.7}\end{equation}
where
\begin{equation}\label{3.8}
 \langle f(x), g(y) \rangle = \frac{1}{2} \sum_{y=-\infty}^{\infty} 
\left[ \sum_{x=-\infty}^{y-1} \sqrt{w(x)} f(x) - \sum_{x=y+1}^{\infty} 
\sqrt{w(x)} f(x) \right] \sqrt{w(y)} g(y).
\end{equation}
\par
When the weight function $w(x)$ is given by eq. (\ref{1.6}), 
the skew orthogonal 
polynomials $R_n(x)$ are compactly expressed 
in terms of the symmetric Hahn polynomials.
To show this we utilize a method analogous to that employed 
by Nagao and Wadati\cite{NW2} for the continuous analogue of
(\ref{3.8}). 
Let us first note the identity
\begin{eqnarray}
\langle f(x), g(y+1)-g(y) \rangle & = & - \langle f(x), \frac{
\sqrt{w(y)} - \sqrt{w(y-1)}}{\sqrt{w(y)}} g(y) \rangle \nonumber \\ 
& - & \frac{1}{2} 
\sum_{x=-\infty}^{\infty} \sqrt{w(x)} f(x) \left[ 
\sqrt{w(x)} g(x + 1) + \sqrt{w(x-1)} g(x) \right] \nonumber \\ 
\end{eqnarray}
and make substitutions
\begin{eqnarray}
f(x) & = & R_n(x), \nonumber \\ 
g(x) & = & \left[ \frac{L}{2} - x + 1 \right] \left[ \frac{L}{2} + x \right] 
C_l(x - 1).
\end{eqnarray}
Noting 
\begin{equation}
\frac{\sqrt{w(x)} - \sqrt{w(x-1)}}{\sqrt{w(x)}} = \frac{1 - 2 x}{
\displaystyle \frac{L}{2} - x + 1}
\end{equation}
and
\begin{eqnarray}
& & \sqrt{w(x)} g(x + 1) + \sqrt{w(x-1)} g(x) \nonumber \\ 
& = & \sqrt{w(x)} \left[ \frac{L}{2} - x \right] \left[ 
\frac{L}{2} + x + 1 \right] C_l(x) + 
\sqrt{w(x)} \left[ 
\frac{L}{2} + x \right]^2 C_l(x-1) \nonumber \\   
& = & \sqrt{w(x)} [ L - 1 + \alpha_l ] C_{l+1}(x) + 
\sqrt{w(x)} \left[ \frac{L(L+1)}{2} + \beta_l \right] C_l(x) 
\nonumber \\ & + & \sqrt{w(x)} [(L-1) \omega_l + 
\gamma_l ] C_{l-1}(x), \nonumber \\ 
\end{eqnarray}
we obtain 
\begin{eqnarray}\label{3.13}
& & \langle R_n(x), \left[ \frac{L}{2} - y \right] \left[ 
\frac{L}{2} + y + 1 \right] C_l(y) \rangle - 
\langle R_n(x), \left[ \frac{L}{2} + y \right]^2 
C_l(y-1) \rangle \nonumber \\ 
& = & - \frac{1}{2} \sum_{x=-\infty}^{\infty} w(x) R_n(x) 
(L- l - 1) C_{l+1}(x)   
 - \frac{1}{4} \sum_{x=-\infty}^{\infty} w(x) R_n(x) 
(L- l + 1) (L-l) C_l(x)  \nonumber \\   
& + & \frac{1}{8} \sum_{x=-\infty}^{\infty} w(x) R_n(x) 
\frac{l (2 L - l + 2) (L - l + 1)^2 (L - l 
+ 2)}{(2 L - 2 l + 3) (2 L - 2 l + 1)} C_{l-1}(x).
\end{eqnarray}  
Putting the expansion
\begin{eqnarray}
R_{2 m}(x) & = & \sum_{j=0}^{m} \alpha_{2 m \ 2 j} C_{2 j}(x), \nonumber \\  
R_{2 m + 1}(x) & = & \sum_{j=0}^{m} \alpha_{2 m + 1 \ 2 j + 1} C_{2 j + 1}(x)
\end{eqnarray}
($\alpha_{jj} = 1$) into (\ref{3.13}) 
and using eqs. (\ref{2.7}) and (\ref{3.7}), we arrive at  
\begin{eqnarray}\label{3.15}
R_{2 m}(x) & = & C_{2 m}(x), \nonumber \\  
R_{2 m + 1}(x) & = & 
C_{2 m + 1}(x) - \frac{L - 2 m}{L - 2 m + 1} \frac{h_{2 m}}{
h_{2 m - 1}} C_{2 m - 1}(x)
\end{eqnarray}
and
\begin{equation}
r_m = \frac{1}{4} (L - 2 m) h_{2 m},
\end{equation}
where $h_n$ is given by eq.(\ref{2.8}).

\subsection{The Case $N$ Even}
Let us first consider the case $N$ is even ($N = 2 \nu$). 
Independent of the functional form of $w(x)$,
the number 
of paths has a quaternion determinant expression \cite{MAHOUX}
\begin{equation}
P_1(x_1,x_2,\cdots,x_N) = C_{MN} 2^{\nu} (\prod_{n=0}^{\nu - 1} 
r_n)  {\rm Tdet}[f(x_j,x_l)]_{j,l=1,2,\cdots,2 \nu}, 
\end{equation}
where the quaternion elements are represented as
\begin{equation}
f(x,y) = \left[ \begin{array}{cc} S(x,y) & I(x,y) \\ 
D(x,y) & S(y,x) \end{array} \right]
\end{equation}
with
\begin{eqnarray}\label{3.19}
S(x,y) & = & \sqrt{w(y)} \sum_{n=0}^{(N/2)-1} \frac{1}{r_n} [\Phi_{2 n}(x) 
R_{2 n + 1}(y) - \Phi_{2 n + 1}(x) R_{2 n}(y) ], \nonumber \\ 
D(x,y) & = & \sqrt{w(x) w(y)} \sum_{n=0}^{(N/2)-1} \frac{1}{r_n} [R_{2 n}(x) 
R_{2 n + 1}(y) - R_{2 n + 1}(x) R_{2 n}(y) ], \nonumber \\ 
I(x,y) & = & - \sum_{n=0}^{(N/2)-1} \frac{1}{r_n} [\Phi_{2 n}(x) 
\Phi_{2 n + 1}(y) - \Phi_{2 n + 1}(x) \Phi_{2 n}(y) ] - \epsilon(x-y)
\end{eqnarray} 
and
\begin{equation}
\Phi_n(x) = \sum_{y=-\infty}^{\infty} \epsilon(x - y) \sqrt{w(y)} R_n(y).
\end{equation}
Here $\epsilon(x)$ is defined as
\begin{equation}
\epsilon(x) = \left\{ \begin{array}{l} 1/2, \ \ x > 0, \\ 
0, \ \ x=0, \\ -1/2, \ \ x < 0. \end{array} \right.
\end{equation}
Using the skew orthogonality relation (\ref{3.7}), 
we can prove the following. 
Let us define
\begin{equation}\label{3.22}
Q_n(x_1,x_2,\cdots,x_n) = {\rm Tdet}[f(x_j,x_l)]_{j,l=1,2,\cdots,n}.
\end{equation}
Then
\begin{equation}\label{3.23}
\sum_{x_n=-\infty}^{\infty} Q_n(x_1,x_2,\cdots,x_n) = (N-n+1) Q_{n-1}(
x_1,x_2,\cdots,x_{n-1}).
\end{equation}
Recursive use of (\ref{3.23}) on (\ref{3.22}) leads to  
\begin{equation}
I^{(1)}_k(x_1,x_2,\cdots,x_k) = C_{MN} 2^{\nu} (\prod_{n=0}^{\nu - 1} 
r_n)  {\rm Tdet}[f(x_j,x_l)]_{j,l=1,2,\cdots,k}.
\end{equation}
Thus the correlation functions are expressed in 
a quaternion determinant form.
\par
In the continuous case, it is known
\cite{AFNV} that for $w(x)$ a classical weight the summations
(\ref{3.19}) can be simplified. Similar simplifications can be
undertaken here. First we note the function $\Phi_n(x)$ can be rewritten as 
\begin{equation}
\Phi_n(x) = \frac{1}{2} 
\sum_{y=-\infty}^{\infty} \sum_{z^{\prime}=-\infty}^{\infty} 
\sum_{z=-\infty}^{z^{\prime}-1} [\delta_{yz} \delta_{xz^{\prime}} 
- \delta_{xz} \delta_{yz^{\prime}}] \sqrt{w(y)} R_n(y).
\end{equation}
We put the dual orthogonality relation (\ref{2.9}) and an expansion 
\begin{eqnarray}
C_{2 m}(x) & = & \sum_{j=0}^{m} \beta_{2 m \ 2 j} R_{2 j}(x), \nonumber \\  
C_{2 m + 1}(x) & = & \sum_{j=0}^{m} \beta_{2 m + 1 \ 2 j + 1} R_{2 j + 1}(x)
\end{eqnarray}
in the above equation to find
\begin{eqnarray}\label{3.27}
\Phi_{2 n}(x) & = & \sqrt{w(x)} \sum_{p=n}^{[(L-1)/2]}  \frac{C_{2 p + 1}(x)}
{h_{2 p + 1}} \beta_{2 p + 1 \ 2 n + 1} \ r_n, \nonumber \\ 
\Phi_{2 n + 1 }(x) & = & 
- \sqrt{w(x)} \sum_{p=n}^{[L/2]}  \frac{C_{2 p}(x)}
{h_{2 p}} \beta_{2 p \ 2 n} \ r_n.
\end{eqnarray}
Here $[x]$ is the largest integer not exceeding $x$. 
Substituting (\ref{3.27}) into (\ref{3.19}) and using (\ref{3.15}) yields
\begin{eqnarray}
S(x,y) & = & \sqrt{w(x) w(y)} \sum_{n=0}^{N-1} \frac{1}{h_n} C_n(x) 
C_n(y) \nonumber \\ 
& + & \sqrt{w(x) w(y)} \sum_{p=N/2}^{[(L-1)/2]} \sum_{n=0}^{(N/2)-1} 
\frac{C_{2 p +1}(x)}{h_{2 p + 1}} \beta_{2 p + 1 \ 2 n + 1} R_{2 n + 1}(y)
\nonumber \\ & = & \sqrt{w(x) w(y)} \sum_{n=0}^{N-1} \frac{1}{h_n} C_n(x) 
C_n(y) \nonumber \\ 
& + & \sqrt{w(x) w(y)} C_{N-1}(y) \sum_{p=N/2}^{[(L-1)/2]} \beta_{2 p + 1 \ 
N-1} \frac{C_{2 p +1}(x)}{h_{2 p + 1}}.
\end{eqnarray} 
Making use of
\begin{equation}
\Phi_{N-2}(x) = \sqrt{w(x)} \sum_{p=N/2}^{[(L-1)/2]} \beta_{2 p + 1 \ 
N-1} \frac{C_{2 p + 1}(x)}{h_{2 p + 1}} r_{(N/2)-1} + 
\sqrt{w(x)} \frac{C_{N-1}(x)}{h_{N-1}} r_{(N/2)-1}
\end{equation}
and $R_{N-2}(x) = C_{N-2}(x)$, we can easily find
\begin{eqnarray}\label{3.30}
S(x,y) & = & \sqrt{w(x) w(y)} \sum_{n=0}^{N-2} \frac{1}{h_n} 
C_n(x) C_n(y) \nonumber \\ 
& + & \frac{1}{r_{(N/2)-1}} \sqrt{w(y)} C_{N-1}(y) 
\sum_{z=-\infty}^{\infty} \epsilon(x - z) \sqrt{w(z)} C_{N-2}(z).
\end{eqnarray}
Note that the first term on the right hand side is precisely (\ref{1.11}) 
with $N \mapsto N-1$. This gives a compact expression of 
$S(x,y)$ in terms of the symmetric Hahn polynomials. 
\par
It should be also noted that, using eq. (\ref{3.27}), we can derive
\begin{equation}
\epsilon(y-x) = \sum_{n=0}^{[(L-1)/2]} \frac{1}{r_n} [\Phi_{2 n}(x) 
\Phi_{2 n + 1}(y) - \Phi_{2 n + 1}(x) \Phi_{2 n}(y) ],
\end{equation}
which yields 
\begin{equation}
I(x,y) = \sum_{n=N/2}^{[(L-1)/2]} \frac{1}{r_n} [\Phi_{2 n}(x) 
\Phi_{2 n + 1}(y) - \Phi_{2 n + 1}(x) \Phi_{2 n}(y) ].
\end{equation}
\par
To conclude this subsection, let us remark on the "partition function" 
\begin{eqnarray}
I^{(1)}_0 & = & C_{MN} 2^{\nu} (\prod_{n=0}^{\nu-1} r_n ) \nonumber \\ 
& = & \prod_{n=0}^{(N/2)-1} \frac{(2 n)! (2 L - 4 n + 1)! (2 L - 4 n)!}{
2 (2 L - 2 n + 1)! [(L - 2 n)!]^2}.
\end{eqnarray}
It can be readily rewritten as
\begin{equation}
I^{(1)}_0 = \prod_{l=0}^{N-1} \frac{1}{(2 M + 2 l + 1)!} 
\prod_{n=0}^{(N/2)-1} \frac{(2 n)! (2 M + 4 n + 3)! (2 M + 4 n + 2)! 
(2 M + 2 n + 1)!}{2 [(M + 2 n + 1)!]^2},
\end{equation}
so that
\begin{equation}
\frac{\left. I^{(1)}_0 \right|_{N \mapsto N+2}}{
\left. I^{(1)}_0 \right|_{N \mapsto N}} = \frac{N! (2 M + N + 1)!}{
(M + N + 1)! (M + N)!} = \prod_{1 \leq i \leq j \leq M} 
\frac{N + i + j + 1}{N + i + j - 1}.
\end{equation}
Therefore, noting 
\begin{equation}
\left. I^{(1)}_0 \right|_{N \mapsto 2} = \frac{1}{2} 
\frac{(2 M + 2)!}{[(M+1)!]^2} = \prod_{1 \leq i \leq j \leq M} 
\frac{i + j + 1}{i + j - 1},
\end{equation}
we obtain
\begin{equation}
I^{(1)}_0 = 
\prod_{1 \leq i \leq j \leq M} \frac{N + i + j - 1}{i + j - 1}.
\end{equation}
This expression is already known. It was conjectured by Essam and 
Guttmann \cite{EG} and subsequently proved in Ref.\cite{GOV} 
using the correspondence 
between the paths of vicious walkers and Young tableaux. In the 
context of Young tableaux, this enumeration problem is reduced to 
a special case of the result known as Bender-Knuth conjecture. 

\subsection{The Case $N$ Odd}
In order to deal with the case $N$  odd ($N = 2 \nu + 1$), we 
introduce the polynomials $R^{\rm odd}_n(x)$,
\begin{equation}\label{3.38}
R^{\rm odd}_n(x) = R_n(x) - \frac{s_n}{s_{N-1}} R_{N-1}(x), 
\ \ n = 0,1,2,\cdots,N-2,  
\end{equation}
where
\begin{equation}\label{3.39}
s_n = \sum_{x=-\infty}^{\infty} \sqrt{w(x)} R_n(x).
\end{equation}
Then we know from \cite{FP95} that we
can express the number of paths in a quaternion 
determinant form 
\begin{equation}
P_1(x_1,x_2,\cdots,x_N) = C_{MN} \ s_{N-1} \ 2^{\nu} (\prod_{n=0}^{\nu - 1} 
r_n)  {\rm Tdet}[f^{\rm odd}(x_j,x_l)]_{j,l=1,2,\cdots,2 \nu + 1}, 
\end{equation}
where the $2 \times 2$ representations of the quaternion elements are 
given by
\begin{equation}
f^{\rm odd}(x,y) = \left[ \begin{array}{cc} S^{\rm odd}(x,y) & 
I^{\rm odd}(x,y) \\ D^{\rm odd}(x,y) & S^{\rm odd}(y,x) \end{array} \right]
\end{equation}
with
\begin{eqnarray}\label{3.42}
S^{\rm odd}(x,y) & = & 
\sqrt{w(y)} \sum_{n=0}^{(N-3)/2} \frac{1}{r_n} [\Phi^{\rm odd}_{2 n}(x) 
R^{\rm odd}_{2 n + 1}(y) - \Phi^{\rm odd}_{2 n + 1}(x) 
R^{\rm odd}_{2 n}(y) ] \nonumber \\ 
& + & \frac{1}{s_{N-1}} \sqrt{w(y)} R_{N-1}(y), 
\nonumber \\ 
D^{\rm odd}(x,y) & = & \sqrt{w(x) w(y)} \sum_{n=0}^{(N-3)/2} 
\frac{1}{r_n} [R^{\rm odd}_{2 n}(x) 
R^{\rm odd}_{2 n + 1}(y) - R^{\rm odd}_{2 n + 1}(x) 
R^{\rm odd}_{2 n}(y) ], \nonumber \\ 
I^{\rm odd}(x,y) & = & - \sum_{n=0}^{(N-3)/2} 
\frac{1}{r_n} [\Phi^{\rm odd}_{2 n}(x) 
\Phi^{\rm odd}_{2 n + 1}(y) - \Phi^{\rm odd}_{2 n + 1}(x) 
\Phi^{\rm odd}_{2 n}(y) ] \nonumber \\ 
& + & \frac{1}{s_{N-1}} \left[ \Phi_{N-1}(x) - \Phi_{N-1}(y) \right] 
- \epsilon(x-y)
\end{eqnarray} 
and
\begin{equation}
\Phi^{\rm odd}_n(x) = 
\sum_{y=-\infty}^{\infty} \epsilon(x - y) \sqrt{w(y)} R^{\rm odd}_n(y).
\end{equation}
The skew orthogonality relation among $R_n(x)$ enables us to carry 
out the multiple integration one by one as before and arrive at 
\begin{equation}
I^{(1)}_k(x_1,x_2,\cdots,x_k) = C_{MN} \ s_{N-1} \ 2^{\nu} 
(\prod_{n=0}^{\nu - 1} r_n)  {\rm Tdet}[f^{\rm odd}(x_j,x_l)]_{j,
l=1,2,\cdots,k}.
\end{equation}
This is the quaternion determinant formula for the correlation 
functions in the case $N$ odd.
\par
Inserting (\ref{3.38}) into (\ref{3.42}), we can rewrite $S^{\rm odd}(x,y)$ as
\begin{eqnarray}
S^{\rm odd}(x,y) & = & \left. S(x,y) \right|_{N \mapsto N-1} 
+ \frac{\sqrt{w(y)}}{s_{N-1}} R_{N-1}(y) \nonumber \\ 
& + & \frac{\sqrt{w(y)}}{s_{N-1}} \sum_{n=0}^{(N-3)/2} \frac{s_{2 n}}{r_n} 
\left[ \Phi_{2 n + 1}(x) R_{N-1}(y) - \Phi_{N-1}(x) R_{2 n + 1}(y) \right].
\end{eqnarray}
If we note from eqs.(\ref{3.19}) and (\ref{3.30}) that
\begin{eqnarray}
\sqrt{w(y)} \sum_{n=0}^{(N-3)/2} \frac{s_{2 n}}{r_n} R_{2 n + 1}(y)  
& = & 2 \lim_{x \rightarrow \infty} \left. S(x,y) 
\right|_{N \mapsto N-1} \nonumber \\ 
& = & \sqrt{w(y)} \frac{s_{N-3}}{r_{(N-3)/2}} C_{N-2}(y), 
\end{eqnarray}
this can be further simplified as
\begin{eqnarray}
S^{\rm odd}(x,y) & = & \left. S(x,y) \right|_{N \mapsto N-1} 
+ \frac{\sqrt{w(y)}}{s_{N-1}} C_{N-1}(y) \nonumber \\ 
& + & \sqrt{w(y)} \frac{s_{N-3}}{s_{N-1} \ r_{(N-3)/2}} 
\left[ \varphi_{N-2}(x) C_{N-1}(y) - \varphi_{N-1}(x) C_{N-2}(y) \right],
\end{eqnarray} 
where
\begin{equation}
\varphi_n(x) = \sum_{y=-\infty}^{\infty} \epsilon(x-y) \sqrt{w(y)} C_n(y).
\end{equation}
\par
We remark that $s_n$ as defined by (\ref{3.39}) can be written in a
closed form. For this purpose we require
the generating functions for the dual Hahn polynomials in 
Ref.\cite{RK}. One of them can be written in terms of the 
Hahn polynomials as 
\begin{equation}
(1-t)^{L-n} \ _2F_1(-n,-n - b;a+1;t) = \sum_{x=0}^L 
\frac{(-L)_x}{x!} Q_n(x) t^x.
\end{equation}
Putting $t=-1$ and $a=b=-L-1$, we deduce 
\begin{equation}\label{3.50}
s_n = \left\{ \begin{array}{ll} \displaystyle 2^{L-n} \frac{n! (L-(n/2))! 
(2 L - 2 n + 1)!}{(n/2)! (2 L - n + 1)! [(L-n)!]^2}, & \ \ n \ {\rm even},
\\ \displaystyle 0, & \ \ n \ {\rm odd}. \end{array} \right.
\end{equation}
As an application,
since the "partition function" $I^{(1)}_0$ can be rewritten as
\begin{eqnarray}
I^{(1)}_0 & = & C_{MN} \ s_{N-1} \ 2^{\nu} (\prod_{n=0}^{\nu-1} r_n ) 
\nonumber \\ & = & s_{N-1} M! \prod_{l=0}^{N-2} \frac{1}{(2 M + 2 l + 3)!} 
\prod_{n=0}^{(N-3)/2} \frac{(2 n)! (2 M + 4 n + 5)! (2 M + 4 n + 4)! 
(2 M + 2 n + 3)!}{2 [(M + 2 n + 2)!]^2}, \nonumber \\ 
\end{eqnarray}
we can utilize the summation formula (\ref{3.50}) to derive
\begin{equation}
\frac{\left. I^{(1)}_0 \right|_{N \mapsto N+2}}{
\left. I^{(1)}_0 \right|_{N \mapsto N}} = \frac{N! (2 M + N + 1)!}{
(M + N + 1)! (M + N)!} = \prod_{1 \leq i \leq j \leq M} 
\frac{N + i + j + 1}{N + i + j - 1},
\end{equation}
which leads to  
\begin{equation}
I^{(1)}_0 = 
\prod_{1 \leq i \leq j \leq M} \frac{N + i + j - 1}{i + j - 1}.
\end{equation}
This is again a known result \cite{GOV},
related to the Bender-Knuth conjecture 
in the theory of Young tableaux.

\section{Continuous Limit}
\setcounter{equation}{0}
\renewcommand{\theequation}{4.\arabic{equation}}
In this last section we discuss the continuous limit to the 
Gaussian ensembles. 
Some aspects of this limit for the $\beta = 2$ Hahn ensemble
(\ref{1.7}) have been discussed in \cite{KJ2}.
Let us first define
\begin{equation}
{\bar w(x)} = \frac{\pi L}{2} \left( \frac{L!}{2^L} \right)^2 
w\left(\frac{\sqrt{L}}{2} x \right).
\end{equation}
Then the Stirling's asymptotic formula for the Gamma function yields
\begin{equation}
w^{(G)}(x) = \lim_{L \rightarrow \infty} {\bar w(x)} = 
{\rm e}^{-x^2}.
\end{equation}
Hence we can say that the vicious walker problem is reduced to 
the Gaussian ensembles of random matrices in this scaling limit, with
(\ref{1.4}) identical to the probability density for the Gaussian
orthogonal ensemble.
\par
We further define monic polynomials
\begin{equation}
{\bar C}_n(x) = \left( \frac{2}{\sqrt{L}} \right)^n C_n\left( 
\frac{\sqrt{L}}{2} x \right)
\end{equation}
and 
\begin{equation}
C^{(G)}_n(x) = \lim_{L \rightarrow \infty} {\bar C}_n(x).
\end{equation}
Then, from eq.(\ref{2.7}), it can be readily seen that
\begin{equation}
\frac{2}{\sqrt{L}} \sum_{x=-\infty}^{\infty} {\bar w}\left(
\frac{2}{\sqrt{L}} x \right) {\bar C}_j\left(\frac{2}{\sqrt{L}} x \right) 
{\bar C}_l\left(\frac{2}{\sqrt{L}} x \right) 
= \frac{\pi L}{2} \left( \frac{L!}{2^L} \right)^2 \left( 
\frac{2}{\sqrt{L}} \right)^{2 j + 1} h_j \delta_{jl}.
\end{equation}
In the asymptotic limit $L \rightarrow \infty$, the infinite sum in the 
above equation becomes a Riemann integral and results in
\begin{equation}
\int_{-\infty}^{\infty} {\rm e}^{- x^2} C^{(G)}_j(x) C^{(G)}_l(x) 
{\rm d}x = \sqrt{\pi} \frac{j!}{2^j} \delta_{jl},
\end{equation}
which means
\begin{equation}
C^{(G)}_n(x) = \frac{1}{2^n} H_n(x),
\end{equation}
where $H_n(x)$ is the Hermite polynomial. The corresponding 
skew orthogonal polynomials are introduced as 
\begin{equation}
{\bar R}_n(x) = \left( \frac{2}{\sqrt{L}} \right)^n R_n\left( 
\frac{\sqrt{L}}{2} x \right)
\end{equation}
and 
\begin{equation}
R^{(G)}_n(x) = \lim_{L \rightarrow \infty} {\bar R}_n(x).
\end{equation}
Then we can similarly find
\begin{displaymath} 
\langle R^{(G)}_{2m}(x), 
R^{(G)}_{2n+1}(y) \rangle_G = 
- \langle R^{(G)}_{2n+1}(x), R^{(G)}_{2m}(y) 
\rangle_G = \sqrt{\pi} \frac{(2 m)!}{2^{2 m}} 
\delta_{mn}, \end{displaymath}
\begin{equation} \langle R^{(G)}_{2m}(x), R^{(G)}_{2n}(y) \rangle_G = 0, \quad 
 \langle R^{(G)}_{2m+1}(x), R^{(G)}_{2n+1}(y) \rangle_G = 0,\end{equation}
where
\begin{equation}
 \langle f(x), g(y) \rangle_G = \frac{1}{2} \int_{-\infty}^{\infty} 
{\rm d}y \left[ \int_{-\infty}^y {\rm d}x \ 
{\rm e}^{-x^2/2} f(x) - \int_{y}^{\infty} {\rm d}x \ 
{\rm e}^{-x^2/2} f(x) \right] {\rm e}^{-y^2/2} g(y).
\end{equation}
The scaling limit of eq.(\ref{3.15}) 
\begin{eqnarray}
R^{(G)}_{2 m}(x) & = & C^{(G)}_{2 m}(x), \nonumber \\  
R_{2 m + 1}^{(G)}(x) & = & 
C_{2 m + 1}^{(G)}(x) -  m C^{(G)}_{2 m - 1}(x)
\end{eqnarray}
reproduces the known compact expression of $R^{(G)}_n(x)$ (see
e.g.~\cite{MBOOK,NW2,AFNV}).
\par
 The Gaussian ensembles have been thoroughly studied by many authors and 
asymptotic properties in the limit $N \rightarrow \infty$ are 
well known. In the vicious walker problem, Gaussian results are 
valid only when we first take the limit $M \rightarrow \infty$ in 
the above way and then analyze the large $N$ asymptotic behavior. 
Other scaling limits with respect to $M$ and $N$ are not 
so well understood and should be further investigated in 
future works.

\bibliographystyle{plain}

\end{document}